\documentclass[8.5pt,twoside,twocolumn]{article}
\usepackage[super,sort&compress,comma]{natbib} 
\usepackage[version=3]{mhchem}
\usepackage{times,mathptmx}
\usepackage{sectsty}
\usepackage{balance} 

\usepackage{graphicx} 
\usepackage{lastpage}
\usepackage[format=plain,justification=raggedright,singlelinecheck=false,font=small,labelfont=bf,labelsep=space]{caption} 
\usepackage{fancyhdr}
\usepackage{units}
\usepackage{todonotes}
\usepackage{psfrag}

\newcommand{\ie}{{\it i.e.}}
\newcommand{\eg}{{\it e.g.}}

\newcommand{\etal}{{\it et al.}}
\newcommand{\changed}{}

\begin{document}

\thispagestyle{plain}
\fancypagestyle{plain}{
\fancyhead[L]{\includegraphics[height=8pt]{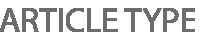}}
\fancyhead[C]{\hspace{-1cm}\includegraphics[height=20pt]{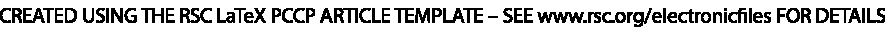}}
\fancyhead[R]{\includegraphics[height=10pt]{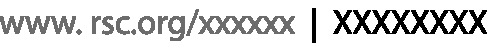}\vspace{-0.2cm}}
\renewcommand{\headrulewidth}{1pt}}
\renewcommand{\thefootnote}{\fnsymbol{footnote}}
\renewcommand\footnoterule{\vspace*{1pt}%
\hrule width 3.4in height 0.4pt \vspace*{5pt}} 
\setcounter{secnumdepth}{5}

\makeatletter 
\def\subsubsection{\@startsection{subsubsection}{3}{10pt}{-1.25ex plus -1ex minus -.1ex}{0ex plus 0ex}{\normalsize\bf}} 
\def\paragraph{\@startsection{paragraph}{4}{10pt}{-1.25ex plus -1ex minus -.1ex}{0ex plus 0ex}{\normalsize\textit}} 
\renewcommand\@biblabel[1]{#1}            
\renewcommand\@makefntext[1]%
{\noindent\makebox[0pt][r]{\@thefnmark\,}#1}
\makeatother 
\renewcommand{\figurename}{\small{Fig.}~}
\sectionfont{\large}
\subsectionfont{\normalsize} 

\fancyfoot{}
\fancyfoot[LO,RE]{\vspace{-7pt}\includegraphics[height=9pt]{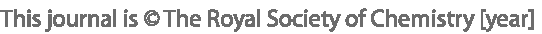}}
\fancyfoot[CO]{\vspace{-7.2pt}\hspace{12.2cm}\includegraphics{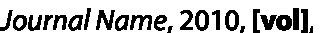}}
\fancyfoot[CE]{\vspace{-7.5pt}\hspace{-13.5cm}\includegraphics{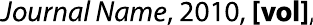}}
\fancyfoot[RO]{\footnotesize{\sffamily{1--\pageref{LastPage} ~\textbar  \hspace{2pt}\thepage}}}
\fancyfoot[LE]{\footnotesize{\sffamily{\thepage~\textbar\hspace{3.45cm} 1--\pageref{LastPage}}}}
\fancyhead{}
\renewcommand{\headrulewidth}{1pt} 
\renewcommand{\footrulewidth}{1pt}
\setlength{\arrayrulewidth}{1pt}
\setlength{\columnsep}{6.5mm}
\setlength\bibsep{1pt}

\twocolumn[
  \begin{@twocolumnfalse}
\noindent\LARGE{\textbf{Monolayer Spontaneous Curvature of Raft-Forming Membrane Lipids$^\dag$}}
\vspace{0.6cm}

\noindent\large{\textbf{Benjamin Kollmitzer,\textit{$^{a}$} Peter Heftberger,\textit{$^{a}$} Michael Rappolt,\textit{$^{b,c}$} and
Georg Pabst$^{\ast}$\textit{$^{a}$}}}\vspace{0.5cm}

\noindent\textit{\small{\textbf{Received Xth XXXXXXXXXX 20XX, Accepted Xth XXXXXXXXX 20XX\newline
First published on the web Xth XXXXXXXXXX 200X}}}

\noindent \textbf{\small{DOI: 10.1039/b000000x}}
\vspace{0.6cm}

\noindent \normalsize{Monolayer spontaneous curvatures for cholesterol, DOPE, POPE, DOPC, DPPC, DSPC, POPC, SOPC, and egg sphingomyelin were obtained using small-angle X-ray scattering (SAXS) on inverted hexagonal phases (\ce{H_{II}}). 
Spontaneous curvatures of bilayer forming lipids were estimated by adding controlled amounts to a \ce{H_{II}} forming template following previously established protocols. Spontanous curvatures of both phosphatidylethanolamines and cholesterol were found to be {\changed at least a factor of two} more negative than those of phosphatidylcholines, whose $J_0$ {\changed are closer} to zero. Interestingly, a significant positive $J_0$ value ($\unit[+0.1]{nm^{-1}}$) was retrieved for DPPC at {\changed \unit[25]{$^\circ$C}}. We further determined the temperature dependence of the spontaneous curvatures $J_0(T)$ in the range from 15 to \unit[55]{$^\circ$C}, resulting in a quite narrow distribution of $-1$ to $\unit[-3 \cdot 10^{-3}]{(nm^{\circ}C)^{-1}}$ for {\changed most} investigated lipids. 
The data allowed us to estimate the monolayer spontaneous curvatures of ternary lipid mixtures showing liquid ordered / liquid disordered phase coexistence.
We report spontaneous curvature phase diagrams for DSPC\slash DOPC\slash Chol, DPPC\slash DOPC\slash Chol and SM\slash POPC\slash Chol and discuss effects on protein insertion and line tension.
}
\vspace{0.5cm}
 \end{@twocolumnfalse}
  ]


\section{Introduction}
\footnotetext{\dag~Electronic Supplementary Information (ESI) available: Electron density maps, tieline parameterization, line tension calculations, {\changed miscibilities, and temperature dependence of spontaneous curvature}. See DOI: 10.1039/b000000x/}
\footnotetext{\textit{$^{a}$~Institute of Molecular Biosciences, Biophysics Division, University of Graz, Austria. Fax: +43 316 4120-390; Tel: +43 316 4120-342; E-mail: Georg.Pabst@uni-graz.at}}
\footnotetext{\textit{$^{b}$~Institute of Inorganic Chemistry, Graz University of Technology, Austria.}}
\footnotetext{\textit{$^{c}$~School of Food Science and Nutrition, University of Leeds, United Kingdom.}}
\addtocounter{footnote}{3}
%
%
%
Curvature is an essential ingredient in a cell's life and occurs most visibly during membrane fusion and fission processes, \eg\ exocytosis and endocytosis, or when a cell is attacked by an enveloped virus.\cite{burger_greasing_2000} Such events may be induced by proteins, but are also known to depend  strongly on the molecular properties of the constituent membrane lipids \cite{chernomordik_mechanics_2008}. For instance membrane fusion can take place in the absence of proteins.\cite{tamm_membrane_2003} Ways to introduce membrane curvature by lipids are, \eg\ by uneven amounts of lipids of the same type in the opposing membrane leaflets or by asymmetric distributions of lipids with different molecular shape due to their different intrinsic curvatures.\cite{sheetz_biological_1974, evans_bending_1974, svetina_membrane_1982, svetina_membrane_1989, bozic_role_1992, miao_budding_1994}

In general, lipids with molecular shapes different from cylinders will form monolayers that either curve away or towards the polar\slash apolar interface.\cite{seddon_polymorphism_1995} In planar membranes however, such monolayers are forced into a flat topology, where they lie back-to-back -- in order to avoid energetically unfavorable voids -- leading to significant curvature elastic stress that is stored within the membrane. This elastic stress may have several functional consequences for membranes and can be viewed as a hidden dimension of membrane curvature. Of particular interest is the role of intrinsic\slash spontaneous curvature in coupling to protein function\cite{cantor_lateral_1997, cantor_influence_1999, safran_statistical_1995, brewster_line_2010, lundbaek_membrane_1996, marsh_lateral_2007, boulgaropoulos_lipid_2012, pabst_effect_2009} and in determining the line tension of lipid domains mimicking membrane rafts\cite{kuzmin_line_2005, akimov_lateral_2007}.

Per definition the spontaneous curvature $J_0 = 0$  for cylindrically formed lipids,  $J_0 < 0$ for lipids with tail regions of bigger lateral cross section than the headgroups and {\it vice versa} for $J_0 > 0$. For example, lipids with negative spontaneous curvature are prone to form non-planar structures like inverted hexagonal phases \ce{H_{II}}. More precisely the radius of curvature of an unstressed monolayer at its {\changed neutral} plane equals $1/J_0$.{\changed \cite{leikin_measured_1996,kozlov_determination_2007}} The {\changed neutral} plane is defined as the position, at which {\changed bending and stretching modes are decoupled, \ie\ bending and stretching deformations proceed independently from each other.\cite{kozlov_elastic_1991-1} A second, frequently quoted surface within the monolayer of amphiphiles is the pivotal plane, which occurs where the molecular area does not change upon deformation. Pioneered by the groups of Rand and Gruner during the late 80ies and the 90ies, the position of this surface and consequently the spontaneous curvature at the pivotal plane, $J_{0p}$ has been determined to high accuracy for a couple of membrane lipids \cite{gruner_directly_1986, tate_temperature_1989, rand_membrane_1990, kozlov_bending_1994, rand_structural_1994, leikin_measured_1996, chen_influence_1997, chen_comparative_1998}, for review see.\cite{zimmerberg_how_2005} The basic idea of these experiments is to use \ce{H_{II}} phases, where the lipid monolayers expose their intrinsic curvature within the individual rods and to determine the pivotal plane by bending and compressing the rods either by gravimetric dehydration or application of osmotic pressure, while measuring the crystalline lattice via X-ray scattering. For a limited number of lipids the neutral plane has been estimated from the pivotal surface using area compressibility and bending rigidity data.\cite{leikin_measured_1996, kozlov_elastic_1991, kozlov_elastic_1991-1} 

In the present work we determine $J_0$ under stress-free conditions by locating the neutral plane from electron density maps of \ce{H_{II}} phases. In particular we focus on spontaneous curvature data of lipids which are involved in the formation of membrane rafts.} Such data is especially of need for calculating protein partitioning in diverse lipid environments \cite{cantor_lateral_1997, cantor_influence_1999, safran_statistical_1995, brewster_line_2010, lundbaek_membrane_1996, marsh_lateral_2007, boulgaropoulos_lipid_2012, pabst_effect_2009} or to estimate the line-tension of lipid domains \cite{akimov_lateral_2007, kuzmin_line_2005}. Additionally, the temperature dependence of spontaneous curvature is still barely investigated. We intend to bridge this gap by determining $J_0$ for cholesterol, DOPC, DPPC, DSPC, POPC, SOPC and egg sphingomyelin within a DOPE matrix from 15 to \unit[55]{$^\circ$C} and for POPE at 37 and \unit[55]{$^\circ$C}.



\section{Materials and methods}
\subsection{Sample preparation}
Cholesterol (Chol), 1,2-dioleoyl-{\it sn}-glycero-3-phosphocholine (DOPC), 1,2-dioleoyl-{\it sn}-glycero-3-phosphoethanolamine (DOPE), 1,2-dipalmitoyl-{\it sn}-glycero-3-phosphocholine (DPPC), 1,2-distearoyl-{\it sn}-glycero-3-phosphocholine (DSPC), 1-palmitoyl-2-oleoyl-{\it sn}-glycero-3-phosphocholine (POPC), 1-stearoyl-2-oleoyl-{\it sn}-glycero-3-phosphocholine (SOPC), and chicken egg sphingomyelin (eggSM) were purchased from Avanti Polar Lipids, Inc., Alabaster, AL, USA and used without further purification. 9-{\it cis}-tricosene was obtained from Sigma-Aldrich, Austria. 

After weighing, lipids were dissolved in chloroform\slash methanol 2:1 at a concentration of \unit[10]{mg/ml}. These lipid stock solutions were mixed in glass vials, \unit[12]{wt\%} tricosene was added and the organic solvent was evaporated under a gentle nitrogen stream. To remove remaining solvent, the samples were placed in vacuum overnight. \unit[18]{M$\Omega$/cm} water (UHQ PS, USF Elga, Wycombe, UK) was added at \unit[20]{$\mu$l/mg lipid} and the mixtures with repeated freeze-thaw cycles fully hydrated. The samples were then protected against oxidation with argon, the vials closed and taped, and stored at \unit[4]{$^\circ$C} for 6--7 days until the measurement. 

\subsection{X-ray measurements}
Small-angle X-ray scattering (SAXS) was performed at the Austrian SAXS beamline at ELETTRA, Trieste.\cite{amenitsch_first_1998, bernstorff_high-throughput_1998} A mar300 Image Plate 2D detector from marresearch, Norderstedt, Germany was used covering a $q$-range from \unit[0.2--6.1]{nm$^{-1}$} and calibrated with silver-behenate (\ce{CH_3(CH_2)_{20}–COOAg}) with a d-spacing of \unit[5.838]{nm}\cite{huang_x-ray_1993}. Sample temperatures were controlled with a bath thermostat from Huber, Offenburg, Germany to a precision of \unit[$\pm$ 0.1]{$^\circ$C}. The samples were equilibrated for \unit[10]{min} at given temperatures before exposure. The exposure time was set to \unit[30]{sec}.

\subsection{X-ray data analysis}
Image integration was performed with FIT2D\cite{hammersley_fit2d:_1997, hammersley_two-dimensional_1996} and cross-checked with MATLAB$^{\textrm \textregistered}$\cite{_matlab_2011}. For further data analysis, homemade MATLAB scripts were used and their function verified with FIT2D\cite{hammersley_mfit:_1989}, IDL$^{\textrm \textregistered}$\cite{_idl_????}, and IGOR Pro$^{\textrm \textregistered}$\cite{_igor_2011}. 

Standard procedures were used to determine the lattice parameters and calculate electron-density maps of the \ce{H_{II}} (for further details see S1 of the ESI\dag). In brief, we applied Lorentzians and additive linear background estimators to fit the Bragg peaks. Typically 5--7 peaks were discernible in the patterns, although for higher temperatures and some samples only three or four peaks could be detected. This was considered in the uncertainty estimations. 

The lattice parameter $a$ was determined via the reflection law, taking into account the information from all Lorentzians. Fourier synthesis yielded the electron density $\rho(\vec{r})$ in real-space, with the phasing condition ($+--+++++-$) known from literature for DOPE-rich, fully hydrated \ce{H_{II}} phases.\cite{turner_x-ray_1992,harper_x-ray_2001,rappolt_conformational_2008} Other phase combinations were tested, but yielded electron densities incompatible with the known structure.

\subsection{Spontaneous curvature estimation}
\subsubsection{Finding the neutral plane.~~} \label{sssec:neutral}
Instead of bending {\changed and compressing} lipid monolayers with osmotic pressures to determine the position $R_0=1/J_0$ of the {\changed neutral plane} \cite{leikin_measured_1996}, we applied the following {\changed procedure, assuming that the neutral plane coincides with the glycerol backbone of phospholipids. This assumption is supported by bending/compression experiments, which always found the pivotal plane to be close to the glycerol backbone of lipid molecules, but slightly within the hydrocarbon region \cite{gruner_directly_1986, tate_temperature_1989, rand_membrane_1990, kozlov_bending_1994, rand_structural_1994, leikin_measured_1996, chen_influence_1997, chen_comparative_1998, rappolt_conformational_2008, alley_x-ray_2008}, while the neutral plane was estimated to be closer to the backbone.\cite{leikin_measured_1996,kozlov_elastic_1991} The proximity of both surfaces to the backbone can be rationalized by the high rigidity in this region.\cite{kozlov_determination_2007} In general, the positions of the neutral and pivotal planes differ by less than 10\% and can even coincide when monolayers are bent in the absence of compression.\cite{leikin_measured_1996, kozlov_determination_2007}

We first locate the  position $R_p$ of the lipid headgroup by fitting a Gaussian to a radial section of the electron density map in a region of $\sim$ \unit[1]{nm} around the maximum value  (see S1 in the ESI\dag\ for further details). Then, the neutral surface is simply given by  $R_0 = R_p + d_{H1}$, where $d_{H1}$ is the distance between the headgroup and the glycerol backbone. Using a joint refinement of X-ray and neutron data on lamellar phases, Ku\v{c}erka and coworkers reported high-resolution structural data for a series of phospholipids.\cite{kucerka_closer_2006, kucerka_structure_2006, kucerka_lipid_2008, kucerka_fluid_2011} The reported $d_{H1}$ range between 0.37 and \unit[0.50]{nm} at temperatures from 20 to \unit[50]{$^\circ$C}. We apply the average of these values for our $R_0$ calculations $d_{H1} = \unit[(0.44 \pm 0.05)]{nm}$. To test the applicability of this procedure, we compare $J_0 = \unit[(-0.387 \pm 0.011)]{nm^{-1}}$ retrieved from the present analysis for DOPE at \unit[25]{$^\circ$C} with  $J_0 = \unit[(-0.367 \pm 0.010)]{nm^{-1}}$ estimated from measurements of the pivotal surface \cite{leikin_measured_1996}. The small difference is expected due to the presence of tricosene in the present experiments in order to reduce packing frustration (see \ref{sssec:packing_frustration}) as compared to the measurements performed by Leikin \etal\cite{leikin_measured_1996}

We also attempted to derive $J_0$ from the width $\sigma_p$ of the Gaussian fitted to the headgroup region of the radial electron density profiles, \ie\ $R_0 = R_p + \sigma_p$. However, the resolution of the electron density maps was for several lipid mixtures too low, yielding  $\sigma_p > 0.7$ nm and hence unrealistic locations of the glycerol backbone.
}

\subsubsection{Relaxation of hexagonal packing frustration.~~} \label{sssec:packing_frustration}
Stress free monolayers, which are necessary for measuring monolayer spontaneous curvature $J_0$, are usually obtained by adding free alkanes or alkenes to inverted hexagonal phases \ce{H_{II}}.\cite{kirk_lyotropic_1985, rand_membrane_1990, chen_comparative_1998, vacklin_bending_2000} By taking up the interstitial spaces, they can reduce the frustration of packing circular objects in a hexagonal manner. This effect is impressively seen for POPE, which forms in the absence of any additive a \ce{H_{II}} phase only above \unit[74]{$^{\circ}$C}.\cite{rappolt_mechanism_2003} Addition of tricosene reduced the frustration to such an amount, that already at \unit[37]{$^{\circ}$C} the \ce{H_{II}} phase was preferred. The total tricosene content of all our samples was \unit[12]{wt\%}. The value was obtained from a test series of varying tricosene concentrations and is close to the \unit[10]{wt\%} used in \cite{alley_x-ray_2008}.

\subsubsection{Spontaneous curvature of  bilayer-forming lipids.~~}
\begin{figure}[h]
\centering
  \includegraphics{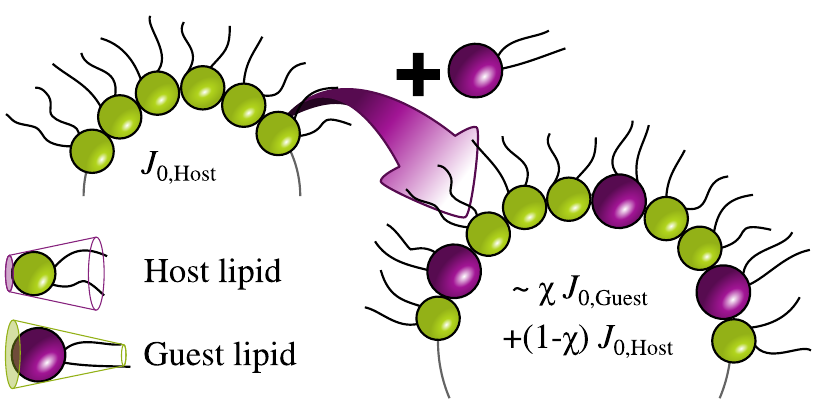}
  \caption{Guest lipid is incorporated at a concentration $\chi$ within the host's template phase. Note the change of the curvature upon mixing.}
  \label{fgr:hostGuestSystem}
\end{figure}

Because monolayer $J_0$ is not accessible in bilayers due to symmetry {\changed constraints}, bilayer-forming lipids have to be incorporated in other structures, see Fig.~\ref{fgr:hostGuestSystem}. Usually \ce{H_{II}} phases (we use the \ce{H_{II}} forming lipid DOPE) are used as templates by mixing the lipid of interest (``guest'') with a \ce{H_{II}}-forming ``host'' lipid.\cite{rand_membrane_1990, chen_influence_1997, kooijman_spontaneous_2005, alley_x-ray_2008, boulgaropoulos_lipid_2012} As long as both lipids mix well, the guest lipid can be expected to modify the curvature of the mixture linearly with respect to its concentration $\chi$ \cite{safran_theory_1990,kozlov_effects_1992,keller_probability_1993,khelashvili_modeling_2009} 
\begin{equation} \label{eq:J0_linear_additivity}
 J_0^{mix} = \chi J_0^{guest} + (1-\chi) J_0^{host},
\end{equation}
and extrapolation towards \unit[100]{\%} gives the spontaneous curvature of the guest lipid\cite{leikin_measured_1996}. A more sophisticated description of spontaneous curvature calculations for lipid mixtures has been reported.\cite{may_spontaneous_1995} However, the experimental determination of several model parameters in this theory remains unclear and experiments seem to contradict with these calculations \cite{gradzielski_droplet_1997}.

\begin{figure}[h]
\centering
  \includegraphics{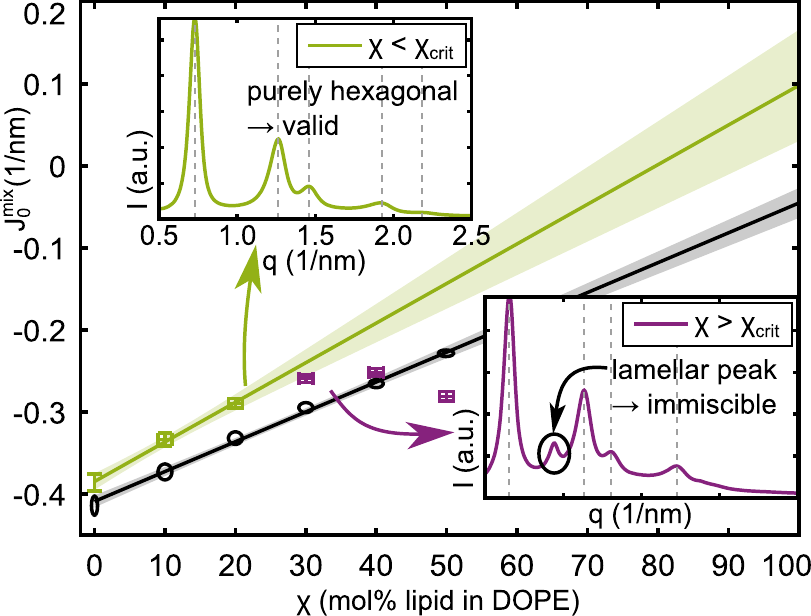}
  \caption{{\changed Determination of $J_0^{DPPC}$ at $\unit[25]{^{\circ}C}$ (crosses) and $J_0^{POPC}$ at $\unit[45]{^{\circ}C}$ (ellipses) by extrapolation of $J_0^{mix}$ towards $\chi = \unit[100]{\%}$. The insets show X-ray patterns for the last valid (top left) and the first immiscible DPPC data point (bottom right).}}
  \label{fgr:J0extrapolation}
\end{figure}

All bilayer-forming lipids were measured at concentrations of 10, 20, 30, 40 and \unit[50]{mol\%} in DOPE. The extrapolation according to Eq.~\eqref{eq:J0_linear_additivity} was performed using all concentrations below a critical value $\chi_{crit}$, at which 
\begin{itemize}
 \item immiscibility was directly observed because non-hexagonal Bragg peaks were visible,
 \item Eq.~\eqref{eq:J0_linear_additivity} did obviously not hold anymore, or
 \item the lattice parameter $a$ did not change smoothly with $\chi$.
\end{itemize}
Entropic contributions get more pronounced at higher temperatures, which generally leads to improved miscibilities. Accordingly, we observed a monotonic increase of $\chi_{crit}$ with $T$ for all samples. An example for the occurrence of non-hexagonal peaks is given in Fig.~\ref{fgr:J0extrapolation}. 

{\changed Good miscibility was observed for Chol and all unsaturated lipids. For saturated lipids $\chi_{crit}$ was not equally satisfactory, but improved above the melting transition of the pure lipid component with the execption of eggSM, where only \unit[10]{mol\%} could be incorporated in the DOPE matrix at alle temperatures. The number of useful data points (where $\chi < \chi_{crit}$) is taken into account for determining the uncertainty of the resulting $J_0$. Extrapolation plots and $\chi_{crit}(T)$ for all lipids are reported in S4 of the ESI\dag.}

\subsubsection{Temperature dependence.~~}
We performed synchrotron SAXS measurements at \unit[10]{$^\circ$C} intervals from \unit[15--55]{$^\circ$C} for all lipids except POPE to quantify the spontaneous curvature's temperature dependence $J_0(T)$. The results could be well described within experimental error by a straight line
\begin{align} 
 J_0(T) &= k \, (T-T^m) + J_0^m \label{eq:J0_of_T} \\
 \Delta J_0(T) &= \sqrt{(\Delta k)^2 \, (T-T^m)^2 + (\Delta J_0^m)^2}, \label{eq:UJ0_of_T}
\end{align}
where we introduced a mean temperature $T^m = \unit[35]{^\circ C}$, the coefficient of thermal curvature change $k$, and $J_0^m$ the spontaneous curvature at $T^m$, while $\Delta X$ denotes the uncertainty of the quantity $X$. POPE was measured at 37 and \unit[55]{$^\circ$C}. Note that POPE forms a \ce{H_{II}} phase at these temperatures only in the presence of an agent such as tricosene, that relaxes the packing frustration. {\changed Fits of $J_0(T)$ in comparison to literature data are plotted in S5 of the ESI\dag.}



\section{Results}

Chol, DOPC, DPPC, DSPC, POPC, SOPC and eggSM were mixed with DOPE and measured as detailed in the previous section. The pure lipids' monolayer spontaneous curvatures for each temperature were obtained by Eq.~\eqref{eq:J0_linear_additivity} {\changed (data in S4 of the ESI\dag)}. Linear fits of the temperature dependence of $J_0$ yielded the values listed in Tab.~\ref{tbl:results_J0} {\changed (fits in S5 of the ESI\dag)}. By inserting these parameters in Eqs.~\eqref{eq:J0_of_T} and \eqref{eq:UJ0_of_T}, $J_0$ and its uncertainty are readily available for any temperature from 15 to \unit[55]{$^\circ$C}. 

POPE was measured with \unit[12]{wt\%} tricosene and excess water at 37 and \unit[55]{$^\circ$C} in the absence of DOPE. Slope and offset of a straight line through the two points following Eq.~\eqref{eq:J0_of_T} with $T^m = \unit[37]{^\circ C}$, are given in Tab.~\ref{tbl:results_J0}. 

\begin{figure}[h]
\centering
  \includegraphics{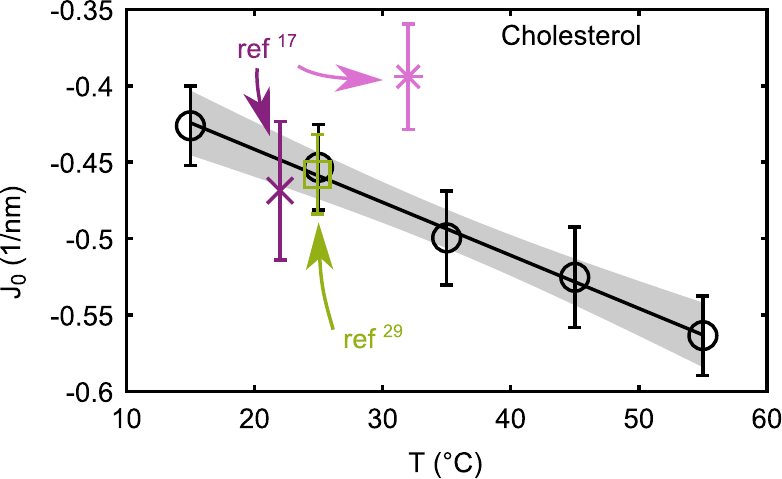}
  \caption{{\changed Comparison between cholesterol spontaneous curvature from literature (refs \cite{chen_influence_1997, boulgaropoulos_lipid_2012})
and new data (circles). Straight line corresponds to linear fit. Literature data at \unit[32]{$^\circ$C} has been determined in a DOPC host matrix, the other two in DOPE.}}
  \label{fgr:CholJ0}
\end{figure}

Figure~\ref{fgr:CholJ0} compares our results for cholesterol with literature data.\footnote{{\changed Reported values for $J_{0p}$ \cite{chen_influence_1997, boulgaropoulos_lipid_2012} were rescaled to $J_0$ using $J_0 \sim J_{0p} (1+\beta)$, with $\beta = 0.065 \pm 0.035$ determined in \cite{leikin_measured_1996}. Data reported by Boulgaropoulos \etal\cite{boulgaropoulos_lipid_2012} were additionally corrected from $J_{0p}= \unit[-0.38]{nm^{-1}}$ to $\unit[-0.43]{nm^{-1}}$ prior to the scaling due to a flaw in their data analysis.}} Although it seems like the literature data has a positive slope of $J_0(T)$, this is probably a coincidence and due to the uncorrelated experiments {\changed in different lipid host systems}. Generally, one would expect the chains to be more flexible and therefore occupy also more space at higher temperature, corresponding to a more negative spontaneous curvature. {\changed This behavior corresponds to $k < 0$, which is the case for all lipids except for eggSM. Most likely this is an artifact due to the limited miscibility of eggSM with DOPE.  Limited miscibility affected also other saturated lipids leading to significant experimental uncertainties in $k$.  Overall $k$ varied in a quite narrow window from $-1$ to $\unit[-3.5\cdot 10^{-3}]{(nm ^\circ C)^{-1}}$, {\it cf.}\ Tab.~\ref{tbl:results_J0}, in good agreement with $k = (-1.7 \pm 0.3)\cdot \unit{10^{-3} (nm ^\circ C)^{-1}}$, reported for DOPE at temperatures from $15$ to $\unit[30]{^\circ C}$.}\cite{kozlov_bending_1994}

\begin{table}[h]
\small
  \caption{\ {\changed Parameters describing $J_0(T)$ according to Eqs.~\eqref{eq:J0_of_T} and \eqref{eq:UJ0_of_T} with $T^m = \unit[35]{^\circ C}$, except (*) where $T^m = \unit[37]{^\circ C}$}}
  \label{tbl:results_J0}
  \begin{tabular*}{0.5\textwidth}{@{\extracolsep{\fill}}lll}
    \hline
    lipid & $J_0^m \pm \Delta J_0^m \, (\unit{1/nm})$ & $k \pm \Delta k \, (\unit{10^{-3}/nm^{\circ}C})$\\
    \hline
   DOPE     & $-0.399 \, \pm \, 0.005$ & $-1.3 \, \pm \, 0.4$\\
   POPE (*) & $-0.316 \, \pm \, 0.007$ & $-2.7 \, \pm \, 0.7$\\
   Chol     & $-0.494 \, \pm \, 0.013$ & $-3.5 \, \pm \, 0.9$\\
   DOPC     & $-0.091 \, \pm \, 0.008$ & $-1.1 \, \pm \, 0.6$\\
   DPPC     & $+0.068 \, \pm \, 0.032$ & $-3.5 \, \pm \, 2.3$\\
   DSPC     & $-0.100 \, \pm \, 0.044$ & $-0.2 \, \pm \, 3.4$\\
   POPC     & $-0.022 \, \pm \, 0.010$ & $-1.8 \, \pm \, 0.7$\\
   SOPC     & $-0.010 \, \pm \, 0.018$ & $-2.2 \, \pm \, 1.3$\\
   eggSM    & $-0.134 \, \pm \, 0.072$ & $+1.4 \, \pm \, 5.1$\\
    \hline
  \end{tabular*}
\end{table}

Interestingly, DPPC is the only bilayer-forming lipid with a significant positive $J_0$. DSPC, for example, with the same headgroup but longer chains has {\changed $J_0 = \unit[-0.1]{nm^{-1}}$ at \unit[35]{$^\circ$C}}. Thus, the headgroup contribution to the molecular shape dominates the crossectional area and hence $J_0$ of DPPC, whereas heads and chains contribute about equally for DSPC. Mismatch in lateral areas of head and chain is known to cause chain tilt and the ripple phase for saturated phosphatidylcholines in a certain range of chain lengths.\cite{koynova_phases_1998} Surprisingly, {\changed $J_0 \sim \unit[-0.1]{nm^{-1}}$} also for eggSM, which like PCs has a choline moiety in the headgroup and is predominantly composed of the same hydrocarbons  as DPPC. Here the sphingosine backbone of eggSM seems to make the difference by taking up more lateral space than the glycerol backbone of PCs. A detailed investigation of this effect is, however, beyond the scope of the present work. {\changed }



\section{Discussion} \label{sec:discussion}

\subsection{Monolayer spontaneous curvature of phase separated systems}

For known compositions, monolayer spontaneous curvatures of mixtures are readily computable by generalization of Eq.~\eqref{eq:J0_linear_additivity} to more components, resulting in 
\begin{equation}
 \label{eq:J0_linear_additivity_ternary}
 J_0^{mix} = \sum_i \chi_i J_0^{(i)}.
\end{equation}
As already mentioned, miscibility is required for the linear additivity of spontaneous curvatures. We assume that this criterion is fulfilled within individual domains of a phase separated system, \ie\ non-ideal mixing is not considered. Thus if the compositions of coexisting phases are known, Eq.~\eqref{eq:J0_linear_additivity} can be applied to determine their spontaneous curvatures. {\changed In the case of non-ideal mixing, which may occur for example by a preferred location of lipids at the domain boundary, energetic contributions from lipid--lipid interactions and mixing entropies need to be considered (see \eg\ \cite{may_spontaneous_1995}). However, this is beyond the scope of the present paper.}

Compositional phase diagrams including tielines have been published recently for ternary lipid mixtures exhibiting liquid disordered (L$_\textrm{d}$) / liquid ordered (L$_\textrm{o}$) phase coexistence.\cite{uppamoochikkal_orientation_2010, heberle_comparison_2010, ionova_phase_2012} These mixtures are simple lipid-only models for membrane rafts, complex platforms which are thought to enable cellular comunication and material transport.\cite{lingwood_lipid_2010} We parameterized the proposed coexistence regions and tieline fields according to the method introduced by Smith and Freed\cite{smith_determination_2009} and slightly modified by Heberle \etal\cite{heberle_comparison_2010}, whose notation we adopted. Briefly, a given phase coexistence region is approximated via a B\'ezier curve of degree five, while a single variable takes care of the tieline fanning. The parameter $u \in [0,1]$ identifies a particular tieline, with the critical point (tieline of length 0) at $u=0$ and the tieline farthest away from the critical point at $u=1$. More details on this parameterization and the explicit values can be found in S2 of the ESI\dag. 

\begin{figure*}
  \centering
  \includegraphics{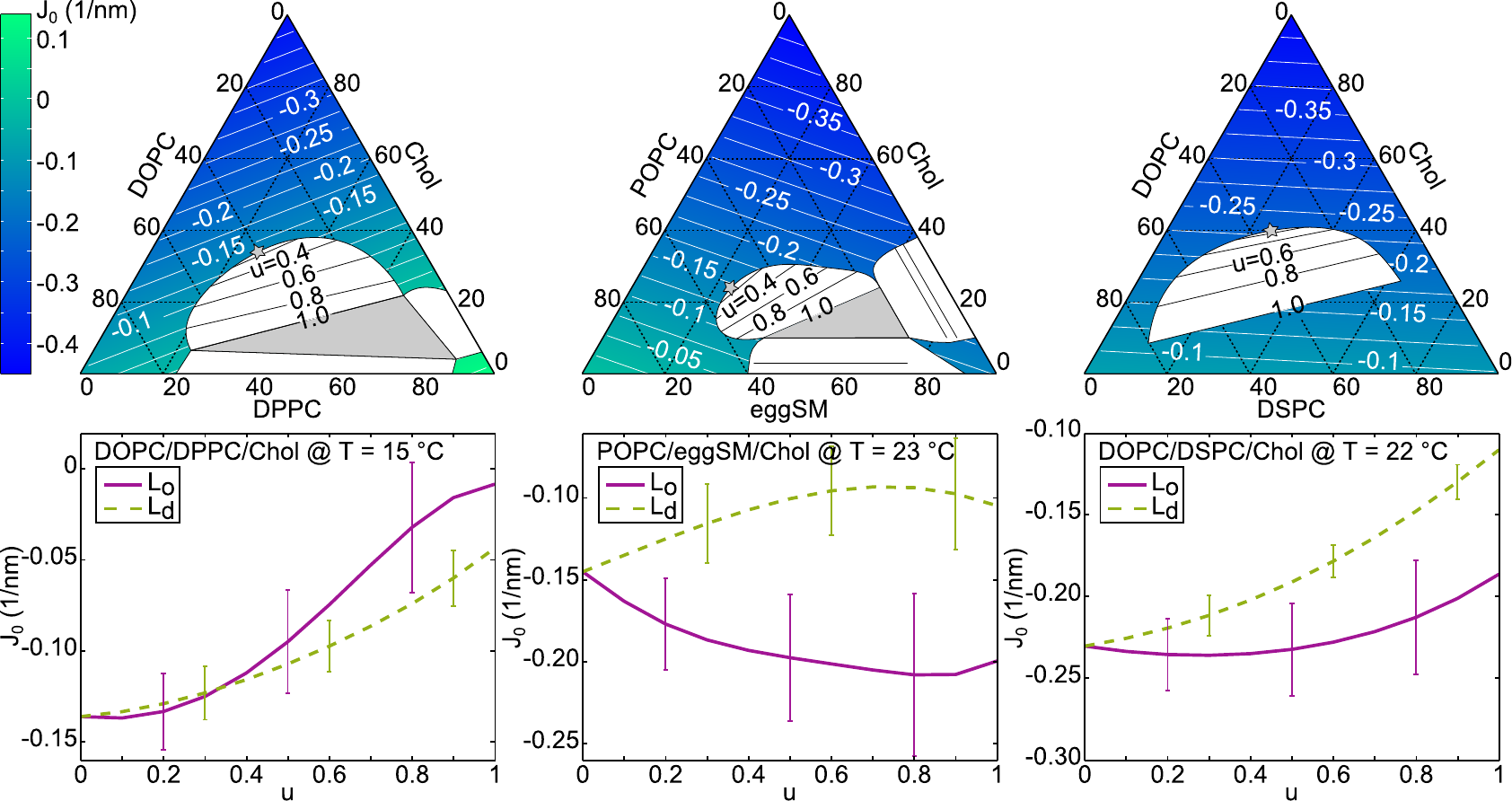}
  \caption{{\changed Spontaneous curvature $J_0$ (white contours and false-color) for three ternary mixtures within the phase diagrams taken from \cite{uppamoochikkal_orientation_2010, heberle_comparison_2010, ionova_phase_2012}. White segments are two-phase coexistence regions with tielines, gray triangles are three-phase coexistence regions, gray stars are critical points (top row). Spontaneous curvature $J_0$ for coexisting L$_\textrm{o}$\slash L$_\textrm{d}$ phases along the boundary of the fluid-fluid phase coexistence regime (bottom row) parameterized by $u$ (see text). }}
  \label{fgr:CompilationJ0vsU}
\end{figure*}

Figure \ref{fgr:CompilationJ0vsU} compares the spontaneous curvatures for coexisting L$_\textrm{o}$\slash L$_\textrm{d}$ phases. The mixture POPC\slash eggSM\slash Chol behaves as expected, \ie\ due to the negative intrinsic curvature of cholesterol, the L$_\textrm{o}$ phase, which contains about twice as much cholesterol as L$_\textrm{d}$ domains, features also a more negative $J_0$. Also DOPC\slash DSPC\slash Chol shows a similar behaviour, although measurement uncertainty limits a clear distinction of the spontaneous curvatures of L$_\textrm{o}$ and L$_\textrm{d}$. For DOPC\slash DPPC\slash Chol however, $J_0$ of the liquid ordered phase at high values of $u$ is less negative than for the L$_\textrm{d}$ phase, and within measurement uncertainty could even be slightly positive. This results from a more positive $J_0$ of DPPC as compared to DSPC with {\changed $J_0 \sim \unit[-0.1]{nm^{-1}}$} (Tab.~\ref{tbl:results_J0}). We note that the quantitative difference between monolayer spontaneous curvatures of L$_\textrm{o}$ and L$_\textrm{d}$ depends on the exact location of the coexistence region and the tieline orientation, which both contain some uncertainties. 

\begin{figure}[h]
\centering
  \includegraphics{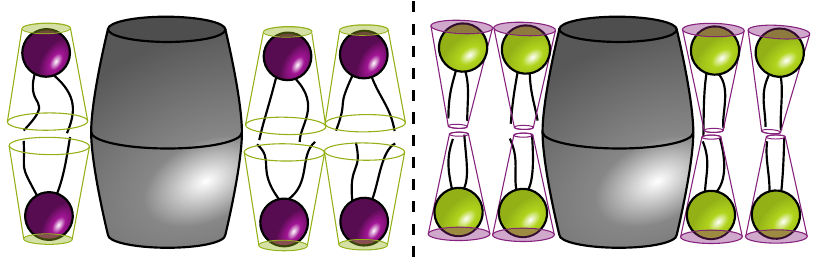}
  \caption{Barrel shaped transmembrane protein within a bilayer composed of lipids with negative (left) and positive (right) monolayer spontaneous curvature. Protein shape reduces the packing frustration within the bilayer in the right case.}
  \label{fgr:LipidProteinInteraction}
\end{figure}

It is instructive to consider the effects of these $J_0$ differences on the insertion probability of simple membrane proteins. Barrel-like transmembrane proteins, which have a thicker cross section at the center of the bilayer than near the bilayer--water interface, would generally prefer phases with positive spontaneous curvature, where the effective lipid cross section at the tail region is smaller than for the headgroup (Fig.~\ref{fgr:LipidProteinInteraction}). In the DOPC\slash DPPC\slash Chol case, this simple argument would mean that the L$_\textrm{o}$ phase is more attractive for such proteins. However already lower-order expansion of the lateral pressure profile reveals a dependence of protein partitioning on further elastic parameters, specifically bending elasticities and Gaussian curvature moduli of L$_\textrm{o}$ and L$_\textrm{d}$.\cite{cantor_lateral_1997, cantor_influence_1999} Literature suggests furthermore hydrophobic mismatch \cite{ben-shaul_molecular_1995} and disturbance of lipid packing \cite{schafer_lipid_2011, domanski_transmembrane_2012} as important factors for determining protein-insertion energies in membranes. Treatment of these effects is beyond the scope of the present work. 

\subsection{Line tension calculation} \label{ssec:line_tension_calculation}

Another parameter that is affected by $J_0$ is the line tension $\gamma$ between two coexisting phases, which influences the size and shape of domains.\cite{garcia-saez_effect_2007, lee_relating_2011} Theory predicts an elastic contribution to $\gamma$ by the monolayer bending moduli, tilt moduli, and thickness difference of L$_\textrm{o}$\slash L$_\textrm{d}$ domains ($\gamma_{el}$) and a second term $\gamma_{J0}$, which includes contributions from the spontaneous curvatures.\cite{kuzmin_line_2005} In the following paragraphs, we give results for the line tension of ternary and quaternary lipid mixtures and discuss the effect of $J_0$. Calculation details, lipid compositions of L$_\textrm{o}$ and L$_\textrm{d}$ phases, as well as elastic parameters are given in S3 of the ESI\dag. {\changed It is important to note, that Helfrich's definition of spontaneous curvature \cite{helfrich_elastic_1973}, which has been applied for deriving $\gamma_{J0}$ in \cite{kuzmin_line_2005}, differs from the quantity $J_0$ we determine in the present work. However, in the case of linear bending behavior, or for small deviations from a flat monolayer, \ie\ if the spontaneous curvature is much smaller than the inverse monolayer thickness $h$, the two values are approximately equal.\cite{kozlov_determination_2007} In S3 of the ESI\dag, we show that indeed $|J_0| < 1/h$ for the following calculations. }

Just recently, bending and tilt moduli, as well as structural parameters have been determined with molecular dynamics (MD) simulations supported by SAXS, for two ternary mixtures showing L$_\textrm{o}$\slash L$_\textrm{d}$ phase separation.\cite{khelashvili_calculating_2013} By combining this information with our new curvature data, we calculate {\changed $\gamma = \unit[1.4]{pN}$} for DOPC\slash DPPC\slash Chol and {\changed $\gamma = \unit[1.6]{pN}$} for DOPC\slash DSPC\slash Chol at given L$_\textrm{o}$\slash L$_\textrm{d}$ compositions. {\changed These values are in the typical range reported from either experiment or theory (see, \eg\cite{risselada_molecular_2008, tian_line_2007, esposito_flicker_2007, honerkamp-smith_line_2008}).} Because of the positive curvature of DPPC, $J_0$ for both phases of DOPC\slash DPPC\slash Chol are close to zero, leading to vanishing contributions of $\gamma_{J0}$ to the line tension. For DOPC\slash DSPC\slash Chol however, the L$_\textrm{o}$ and L$_\textrm{d}$ phase feature a negative $J_0$, leading to {\changed $\gamma_{J0} = \unit[-1.8]{pN}$}, \ie\ the line tension between the coexisting domains is decreased due to the contribution of $J_0$. 

\begin{figure}
\centering
  \includegraphics{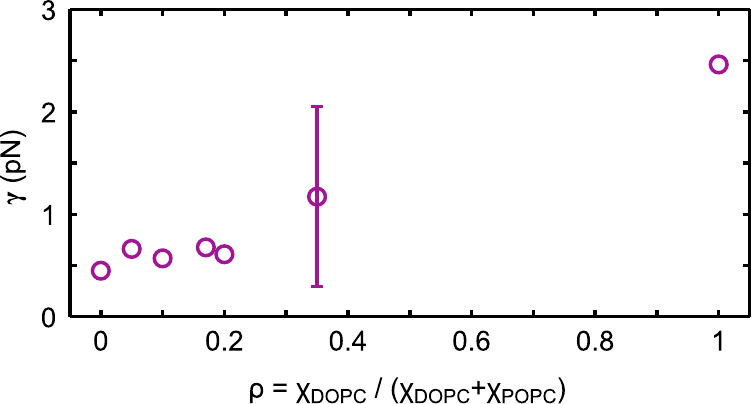}
  \caption{{\changed Calculated line tension $\gamma$ between L$_\textrm{o}$ and L$_\textrm{d}$ domains in DOPC\slash POPC\slash DSPC\slash Chol. Uncertainties of all data points are comparable.}}
  \label{fgr:GammaHeberle}
\end{figure}

The same theory has been applied to rationalize the transition from nanoscopic to microscopic domains, recently reported for the quaternary mixture DOPC\slash POPC\slash DSPC\slash Chol.\cite{heberle_bilayer_2013} Starting from nanometer sized domains in POPC\slash DSPC\slash Chol, replacing POPC with DOPC has lead to increasing domain sizes, and finally to domains in the micrometer regime for DOPC\slash DSPC\slash Chol. Parameterized by the ratio $\rho = \chi_{DOPC}/(\chi_{DOPC}+\chi_{POPC})$, the original calculation of the line tension has explained this behavior, but apart from bilayer thickness information only estimated values for the parameters influencing $\gamma$ were available. Applying bending and tilt moduli from MD simulations \cite{khelashvili_calculating_2013}, spontaneous curvatures from the current work, and structural information from Heberle \etal\cite{heberle_bilayer_2013}, we were able to calculate the line tension for $\rho = 1$ and give improved estimations for $\rho < 1$ (Fig.~\ref{fgr:GammaHeberle}). Because of compositional differences for L$_\textrm{o}$\slash L$_\textrm{d}$ domains between experiment and MD simulation, present calculations still rely on considerable assumptions for $\rho < 1$. In general, the change of nanoscopic to microscopic domains is accompanied by an increase of line tension. This agrees well with our results of {\changed $\gamma \sim \unit[0.5]{pN}$} for the nanoscopic regime, {\changed $\gamma \sim \unit[2.5]{pN}$} for the microscopic regime, and intermediate in between. The contribution of spontaneous curvature to $\gamma$ stays nearly constant for all compositions, meaning the transition from nanoscopic to microscopic domains is mainly driven by bilayer thickness differences in this case, in agreement with the conclusions of the original report \cite{heberle_bilayer_2013}. 

\section{Conclusions}

{\changed Evaluating synchrotron SAXS data of DOPE-rich lipid mixtures in the \ce{H_{II}} phase, we were able to estimate monolayer spontaneous curvatures $J_0$ for several biologically relevant phospholipids, cholesterol and egg sphingomyelin at temperatures ranging from 15 to \unit[55]{$^\circ$C}. Within experimental accuracy, our results are in good agreement with values from more in-depth studies by other groups, conducted at room temperature on DOPE, DOPC, and cholesterol. }

Our measurements extend the $J_0$-list of lipid species and add their temperature dependence.\cite{zimmerberg_how_2005} These data will be useful for numerous applications in membrane biophysics.

In the present work we discuss three examples: (i) the monolayer spontaneous curvatures of raft-like lipid mixtures (ii) line tension of L$_\textrm{o}$\slash L$_\textrm{d}$ phases and (iii) evaluation of the line tension during a transition from nanoscopic to microscopic domains. For the studied mixtures of POPC\slash eggSM\slash Chol and DOPC\slash DSPC\slash Chol, $J_0$ of the L$_\textrm{o}$ phase was found to be more negative than that of the coexisting L$_\textrm{d}$ phase. DOPC\slash DPPC\slash Chol however shows a contrary behavior, with a more positively curved liquid ordered phase due to the positive $J_0$ of DPPC. This would favor partitioning of barrel-shaped proteins into the L$_\textrm{o}$ phase. Regarding line tension, we found only significant contributions of $J_0$ for coexisting domains in DOPC\slash DSPC\slash Chol. In DOPC\slash DPPC\slash Chol and also for the transition from nanoscopic to microscopic domains, $\gamma$ seems to be dominated by elastic moduli and thickness differences.


\section{Acknowledgements}
This work is supported by the Austrian Science Fund FWF, Project No.\ P24459-B20. The authors thank Karl Lohner, George Khelashvili, Siewert-Jan Marrink, and Ilya Levental for valuable discussions {\changed and in particular Daniel Harries for pointing us at literature explaining delicate differences in spontaneous curvatures. }




\footnotesize{

\begin{mcitethebibliography}{77}
\providecommand*{\natexlab}[1]{#1}
\providecommand*{\mciteSetBstSublistMode}[1]{}
\providecommand*{\mciteSetBstMaxWidthForm}[2]{}
\providecommand*{\mciteBstWouldAddEndPuncttrue}
  {\def\EndOfBibitem{\unskip.}}
\providecommand*{\mciteBstWouldAddEndPunctfalse}
  {\let\EndOfBibitem\relax}
\providecommand*{\mciteSetBstMidEndSepPunct}[3]{}
\providecommand*{\mciteSetBstSublistLabelBeginEnd}[3]{}
\providecommand*{\EndOfBibitem}{}
\mciteSetBstSublistMode{f}
\mciteSetBstMaxWidthForm{subitem}
{(\emph{\alph{mcitesubitemcount}})}
\mciteSetBstSublistLabelBeginEnd{\mcitemaxwidthsubitemform\space}
{\relax}{\relax}

\bibitem[Burger(2000)]{burger_greasing_2000}
K.~N. Burger, \emph{Traffic}, 2000, \textbf{1}, 605{\textendash}613\relax
\mciteBstWouldAddEndPuncttrue
\mciteSetBstMidEndSepPunct{\mcitedefaultmidpunct}
{\mcitedefaultendpunct}{\mcitedefaultseppunct}\relax
\EndOfBibitem
\bibitem[Chernomordik and Kozlov(2008)]{chernomordik_mechanics_2008}
L.~V. Chernomordik and M.~M. Kozlov, \emph{Nat. Struct. Mol. Biol.}, 2008,
  \textbf{15}, 675--683\relax
\mciteBstWouldAddEndPuncttrue
\mciteSetBstMidEndSepPunct{\mcitedefaultmidpunct}
{\mcitedefaultendpunct}{\mcitedefaultseppunct}\relax
\EndOfBibitem
\bibitem[Tamm \emph{et~al.}(2003)Tamm, Crane, and
  Kiessling]{tamm_membrane_2003}
L.~K. Tamm, J.~Crane and V.~Kiessling, \emph{Curr. Opin. Struct. Biol.}, 2003,
  \textbf{13}, 453--466\relax
\mciteBstWouldAddEndPuncttrue
\mciteSetBstMidEndSepPunct{\mcitedefaultmidpunct}
{\mcitedefaultendpunct}{\mcitedefaultseppunct}\relax
\EndOfBibitem
\bibitem[Sheetz and Singer(1974)]{sheetz_biological_1974}
M.~P. Sheetz and S.~J. Singer, \emph{{PNAS}}, 1974, \textbf{71},
  4457--4461\relax
\mciteBstWouldAddEndPuncttrue
\mciteSetBstMidEndSepPunct{\mcitedefaultmidpunct}
{\mcitedefaultendpunct}{\mcitedefaultseppunct}\relax
\EndOfBibitem
\bibitem[Evans(1974)]{evans_bending_1974}
E.~Evans, \emph{Biophys. J.}, 1974, \textbf{14}, 923--931\relax
\mciteBstWouldAddEndPuncttrue
\mciteSetBstMidEndSepPunct{\mcitedefaultmidpunct}
{\mcitedefaultendpunct}{\mcitedefaultseppunct}\relax
\EndOfBibitem
\bibitem[Svetina \emph{et~al.}(1982)Svetina, Ottova-Leitmannov{\'a}, and
  Glaser]{svetina_membrane_1982}
S.~Svetina, A.~Ottova-Leitmannov{\'a} and R.~Glaser, \emph{J. Theor. Biol.},
  1982, \textbf{94}, 13--23\relax
\mciteBstWouldAddEndPuncttrue
\mciteSetBstMidEndSepPunct{\mcitedefaultmidpunct}
{\mcitedefaultendpunct}{\mcitedefaultseppunct}\relax
\EndOfBibitem
\bibitem[Svetina and {\v Z}ek{\v s}(1989)]{svetina_membrane_1989}
S.~Svetina and B.~{\v Z}ek{\v s}, \emph{Eur. Biophys. J.}, 1989, \textbf{17},
  101--111\relax
\mciteBstWouldAddEndPuncttrue
\mciteSetBstMidEndSepPunct{\mcitedefaultmidpunct}
{\mcitedefaultendpunct}{\mcitedefaultseppunct}\relax
\EndOfBibitem
\bibitem[Bozic \emph{et~al.}(1992)Bozic, Svetina, Zeks, and
  Waugh]{bozic_role_1992}
B.~Bozic, S.~Svetina, B.~Zeks and R.~E. Waugh, \emph{Biophys. J.}, 1992,
  \textbf{61}, 963--973\relax
\mciteBstWouldAddEndPuncttrue
\mciteSetBstMidEndSepPunct{\mcitedefaultmidpunct}
{\mcitedefaultendpunct}{\mcitedefaultseppunct}\relax
\EndOfBibitem
\bibitem[Miao \emph{et~al.}(1994)Miao, Seifert, Wortis, and
  D{\"o}bereiner]{miao_budding_1994}
L.~Miao, U.~Seifert, M.~Wortis and H.-G. D{\"o}bereiner, \emph{Phys. Rev. E},
  1994, \textbf{49}, 5389--5407\relax
\mciteBstWouldAddEndPuncttrue
\mciteSetBstMidEndSepPunct{\mcitedefaultmidpunct}
{\mcitedefaultendpunct}{\mcitedefaultseppunct}\relax
\EndOfBibitem
\bibitem[Seddon and Templer(1995)]{seddon_polymorphism_1995}
J.~Seddon and R.~Templer, \emph{Handbook of biological physics}, North-Holland,
  1995, vol.~1, p. 97{\textendash}160\relax
\mciteBstWouldAddEndPuncttrue
\mciteSetBstMidEndSepPunct{\mcitedefaultmidpunct}
{\mcitedefaultendpunct}{\mcitedefaultseppunct}\relax
\EndOfBibitem
\bibitem[Cantor(1997)]{cantor_lateral_1997}
R.~S. Cantor, \emph{J. Phys. Chem. B}, 1997, \textbf{101},
  1723{\textendash}1725\relax
\mciteBstWouldAddEndPuncttrue
\mciteSetBstMidEndSepPunct{\mcitedefaultmidpunct}
{\mcitedefaultendpunct}{\mcitedefaultseppunct}\relax
\EndOfBibitem
\bibitem[Cantor(1999)]{cantor_influence_1999}
R.~S. Cantor, \emph{Chem. Phys. Lipids}, 1999, \textbf{101},
  45{\textendash}56\relax
\mciteBstWouldAddEndPuncttrue
\mciteSetBstMidEndSepPunct{\mcitedefaultmidpunct}
{\mcitedefaultendpunct}{\mcitedefaultseppunct}\relax
\EndOfBibitem
\bibitem[Safran(1995)]{safran_statistical_1995}
S.~A. Safran, \emph{J. Stat. Phys.}, 1995, \textbf{78}, 1175--1177\relax
\mciteBstWouldAddEndPuncttrue
\mciteSetBstMidEndSepPunct{\mcitedefaultmidpunct}
{\mcitedefaultendpunct}{\mcitedefaultseppunct}\relax
\EndOfBibitem
\bibitem[Brewster and Safran(2010)]{brewster_line_2010}
R.~Brewster and S.~A. Safran, \emph{Biophys. J.}, 2010, \textbf{98},
  L21--L23\relax
\mciteBstWouldAddEndPuncttrue
\mciteSetBstMidEndSepPunct{\mcitedefaultmidpunct}
{\mcitedefaultendpunct}{\mcitedefaultseppunct}\relax
\EndOfBibitem
\bibitem[Lundb{\ae}k \emph{et~al.}(1996)Lundb{\ae}k, Birn, Girshman, Hansen,
  and Andersen]{lundbaek_membrane_1996}
J.~A. Lundb{\ae}k, P.~Birn, J.~Girshman, A.~J. Hansen and O.~S. Andersen,
  \emph{Biochemistry}, 1996, \textbf{35}, 3825--3830\relax
\mciteBstWouldAddEndPuncttrue
\mciteSetBstMidEndSepPunct{\mcitedefaultmidpunct}
{\mcitedefaultendpunct}{\mcitedefaultseppunct}\relax
\EndOfBibitem
\bibitem[Marsh(2007)]{marsh_lateral_2007}
D.~Marsh, \emph{Biophys. J.}, 2007, \textbf{93}, 3884--3899\relax
\mciteBstWouldAddEndPuncttrue
\mciteSetBstMidEndSepPunct{\mcitedefaultmidpunct}
{\mcitedefaultendpunct}{\mcitedefaultseppunct}\relax
\EndOfBibitem
\bibitem[Boulgaropoulos \emph{et~al.}(2012)Boulgaropoulos, Rappolt, Sartori,
  Amenitsch, and Pabst]{boulgaropoulos_lipid_2012}
B.~Boulgaropoulos, M.~Rappolt, B.~Sartori, H.~Amenitsch and G.~Pabst,
  \emph{Biophys. J.}, 2012, \textbf{102}, 2031--2038\relax
\mciteBstWouldAddEndPuncttrue
\mciteSetBstMidEndSepPunct{\mcitedefaultmidpunct}
{\mcitedefaultendpunct}{\mcitedefaultseppunct}\relax
\EndOfBibitem
\bibitem[Pabst \emph{et~al.}(2009)Pabst, Boulgaropoulos, Gander, Sarangi,
  Amenitsch, Raghunathan, and Laggner]{pabst_effect_2009}
G.~Pabst, B.~Boulgaropoulos, E.~Gander, B.~R. Sarangi, H.~Amenitsch, V.~A.
  Raghunathan and P.~Laggner, \emph{J. Membr. Biol.}, 2009, \textbf{231},
  125--132\relax
\mciteBstWouldAddEndPuncttrue
\mciteSetBstMidEndSepPunct{\mcitedefaultmidpunct}
{\mcitedefaultendpunct}{\mcitedefaultseppunct}\relax
\EndOfBibitem
\bibitem[Kuzmin \emph{et~al.}(2005)Kuzmin, Akimov, Chizmadzhev, Zimmerberg, and
  Cohen]{kuzmin_line_2005}
P.~I. Kuzmin, S.~A. Akimov, Y.~A. Chizmadzhev, J.~Zimmerberg and F.~S. Cohen,
  \emph{Biophys. J.}, 2005, \textbf{88}, 1120--1133\relax
\mciteBstWouldAddEndPuncttrue
\mciteSetBstMidEndSepPunct{\mcitedefaultmidpunct}
{\mcitedefaultendpunct}{\mcitedefaultseppunct}\relax
\EndOfBibitem
\bibitem[Akimov \emph{et~al.}(2007)Akimov, Kuzmin, Zimmerberg, and
  Cohen]{akimov_lateral_2007}
S.~A. Akimov, P.~I. Kuzmin, J.~Zimmerberg and F.~S. Cohen, \emph{Phys. Rev. E},
  2007, \textbf{75}, 011919\relax
\mciteBstWouldAddEndPuncttrue
\mciteSetBstMidEndSepPunct{\mcitedefaultmidpunct}
{\mcitedefaultendpunct}{\mcitedefaultseppunct}\relax
\EndOfBibitem
\bibitem[Leikin \emph{et~al.}(1996)Leikin, Kozlov, Fuller, and
  Rand]{leikin_measured_1996}
S.~Leikin, M.~M. Kozlov, N.~L. Fuller and R.~P. Rand, \emph{Biophys. J.}, 1996,
  \textbf{71}, 2623{\textendash}2632\relax
\mciteBstWouldAddEndPuncttrue
\mciteSetBstMidEndSepPunct{\mcitedefaultmidpunct}
{\mcitedefaultendpunct}{\mcitedefaultseppunct}\relax
\EndOfBibitem
\bibitem[Kozlov(2007)]{kozlov_determination_2007}
M.~M. Kozlov, \emph{Methods in Membrane Lipids}, Springer, 2007, p.
  355{\textendash}366\relax
\mciteBstWouldAddEndPuncttrue
\mciteSetBstMidEndSepPunct{\mcitedefaultmidpunct}
{\mcitedefaultendpunct}{\mcitedefaultseppunct}\relax
\EndOfBibitem
\bibitem[Kozlov and Winterhalter(1991)]{kozlov_elastic_1991-1}
M.~M. Kozlov and M.~Winterhalter, \emph{Journal de Physique {II}}, 1991,
  \textbf{1}, 1077{\textendash}1084\relax
\mciteBstWouldAddEndPuncttrue
\mciteSetBstMidEndSepPunct{\mcitedefaultmidpunct}
{\mcitedefaultendpunct}{\mcitedefaultseppunct}\relax
\EndOfBibitem
\bibitem[Gruner \emph{et~al.}(1986)Gruner, Parsegian, and
  Rand]{gruner_directly_1986}
S.~M. Gruner, V.~A. Parsegian and R.~P. Rand, \emph{Faraday Discuss.}, 1986,
  \textbf{81}, 29--37\relax
\mciteBstWouldAddEndPuncttrue
\mciteSetBstMidEndSepPunct{\mcitedefaultmidpunct}
{\mcitedefaultendpunct}{\mcitedefaultseppunct}\relax
\EndOfBibitem
\bibitem[Tate and Gruner(1989)]{tate_temperature_1989}
M.~W. Tate and S.~M. Gruner, \emph{Biochemistry}, 1989, \textbf{28},
  4245{\textendash}4253\relax
\mciteBstWouldAddEndPuncttrue
\mciteSetBstMidEndSepPunct{\mcitedefaultmidpunct}
{\mcitedefaultendpunct}{\mcitedefaultseppunct}\relax
\EndOfBibitem
\bibitem[Rand \emph{et~al.}(1990)Rand, Fuller, Gruner, and
  Parsegian]{rand_membrane_1990}
R.~P. Rand, N.~L. Fuller, S.~M. Gruner and V.~A. Parsegian,
  \emph{Biochemistry}, 1990, \textbf{29}, 76{\textendash}87\relax
\mciteBstWouldAddEndPuncttrue
\mciteSetBstMidEndSepPunct{\mcitedefaultmidpunct}
{\mcitedefaultendpunct}{\mcitedefaultseppunct}\relax
\EndOfBibitem
\bibitem[Kozlov \emph{et~al.}(1994)Kozlov, Leikin, and
  Rand]{kozlov_bending_1994}
M.~M. Kozlov, S.~Leikin and R.~P. Rand, \emph{Biophys. J.}, 1994, \textbf{67},
  1603--1611\relax
\mciteBstWouldAddEndPuncttrue
\mciteSetBstMidEndSepPunct{\mcitedefaultmidpunct}
{\mcitedefaultendpunct}{\mcitedefaultseppunct}\relax
\EndOfBibitem
\bibitem[Rand and Fuller(1994)]{rand_structural_1994}
R.~P. Rand and N.~L. Fuller, \emph{Biophys. J.}, 1994, \textbf{66},
  2127{\textendash}2138\relax
\mciteBstWouldAddEndPuncttrue
\mciteSetBstMidEndSepPunct{\mcitedefaultmidpunct}
{\mcitedefaultendpunct}{\mcitedefaultseppunct}\relax
\EndOfBibitem
\bibitem[Chen and Rand(1997)]{chen_influence_1997}
Z.~Chen and R.~P. Rand, \emph{Biophys. J.}, 1997, \textbf{73}, 267--276\relax
\mciteBstWouldAddEndPuncttrue
\mciteSetBstMidEndSepPunct{\mcitedefaultmidpunct}
{\mcitedefaultendpunct}{\mcitedefaultseppunct}\relax
\EndOfBibitem
\bibitem[Chen and Rand(1998)]{chen_comparative_1998}
Z.~Chen and R.~P. Rand, \emph{Biophys. J.}, 1998, \textbf{74},
  944{\textendash}952\relax
\mciteBstWouldAddEndPuncttrue
\mciteSetBstMidEndSepPunct{\mcitedefaultmidpunct}
{\mcitedefaultendpunct}{\mcitedefaultseppunct}\relax
\EndOfBibitem
\bibitem[Zimmerberg and Kozlov(2005)]{zimmerberg_how_2005}
J.~Zimmerberg and M.~M. Kozlov, \emph{Nat. Rev. Mol. Cell Biol.}, 2005,
  \textbf{7}, 9--19\relax
\mciteBstWouldAddEndPuncttrue
\mciteSetBstMidEndSepPunct{\mcitedefaultmidpunct}
{\mcitedefaultendpunct}{\mcitedefaultseppunct}\relax
\EndOfBibitem
\bibitem[Kozlov and Winterhalter(1991)]{kozlov_elastic_1991}
M.~M. Kozlov and M.~Winterhalter, \emph{Journal de Physique {II}}, 1991,
  \textbf{1}, 1085{\textendash}1100\relax
\mciteBstWouldAddEndPuncttrue
\mciteSetBstMidEndSepPunct{\mcitedefaultmidpunct}
{\mcitedefaultendpunct}{\mcitedefaultseppunct}\relax
\EndOfBibitem
\bibitem[Amenitsch \emph{et~al.}(1998)Amenitsch, Rappolt, Kriechbaum, Mio,
  Laggner, and Bernstorff]{amenitsch_first_1998}
H.~Amenitsch, M.~Rappolt, M.~Kriechbaum, H.~Mio, P.~Laggner and S.~Bernstorff,
  \emph{J. Synchrotron Radiat.}, 1998, \textbf{5}, 506--508\relax
\mciteBstWouldAddEndPuncttrue
\mciteSetBstMidEndSepPunct{\mcitedefaultmidpunct}
{\mcitedefaultendpunct}{\mcitedefaultseppunct}\relax
\EndOfBibitem
\bibitem[Bernstorff \emph{et~al.}(1998)Bernstorff, Amenitsch, and
  Laggner]{bernstorff_high-throughput_1998}
S.~Bernstorff, H.~Amenitsch and P.~Laggner, \emph{J. Synchrotron Radiat.},
  1998, \textbf{5}, 1215--1221\relax
\mciteBstWouldAddEndPuncttrue
\mciteSetBstMidEndSepPunct{\mcitedefaultmidpunct}
{\mcitedefaultendpunct}{\mcitedefaultseppunct}\relax
\EndOfBibitem
\bibitem[Huang \emph{et~al.}(1993)Huang, Toraya, Blanton, and
  Wu]{huang_x-ray_1993}
T.~C. Huang, H.~Toraya, T.~N. Blanton and Y.~Wu, \emph{J. Appl. Crystallogr.},
  1993, \textbf{26}, 180{\textendash}184\relax
\mciteBstWouldAddEndPuncttrue
\mciteSetBstMidEndSepPunct{\mcitedefaultmidpunct}
{\mcitedefaultendpunct}{\mcitedefaultseppunct}\relax
\EndOfBibitem
\bibitem[Hammersley(1997)]{hammersley_fit2d:_1997}
A.~P. Hammersley, \emph{European Synchrotron Radiation Facility Internal Report
  {ESRF97HA02T}}, 1997\relax
\mciteBstWouldAddEndPuncttrue
\mciteSetBstMidEndSepPunct{\mcitedefaultmidpunct}
{\mcitedefaultendpunct}{\mcitedefaultseppunct}\relax
\EndOfBibitem
\bibitem[Hammersley \emph{et~al.}(1996)Hammersley, Svensson, Hanfland, Fitch,
  and Hausermann]{hammersley_two-dimensional_1996}
A.~P. Hammersley, S.~O. Svensson, M.~Hanfland, A.~N. Fitch and D.~Hausermann,
  \emph{High Pressure Res.}, 1996, \textbf{14}, 235--248\relax
\mciteBstWouldAddEndPuncttrue
\mciteSetBstMidEndSepPunct{\mcitedefaultmidpunct}
{\mcitedefaultendpunct}{\mcitedefaultseppunct}\relax
\EndOfBibitem
\bibitem[_ma(2011)]{_matlab_2011}
\emph{{MATLAB} v. 7.12 (R2011a)}, 2011\relax
\mciteBstWouldAddEndPuncttrue
\mciteSetBstMidEndSepPunct{\mcitedefaultmidpunct}
{\mcitedefaultendpunct}{\mcitedefaultseppunct}\relax
\EndOfBibitem
\bibitem[Hammersley and Riekel(1989)]{hammersley_mfit:_1989}
A.~P. Hammersley and C.~Riekel, \emph{Synchrotron Radiation News}, 1989,
  \textbf{2}, 24--26\relax
\mciteBstWouldAddEndPuncttrue
\mciteSetBstMidEndSepPunct{\mcitedefaultmidpunct}
{\mcitedefaultendpunct}{\mcitedefaultseppunct}\relax
\EndOfBibitem
\bibitem[_id()]{_idl_????}
\emph{{IDL} (Interactive Data Language) v. 6.1}\relax
\mciteBstWouldAddEndPuncttrue
\mciteSetBstMidEndSepPunct{\mcitedefaultmidpunct}
{\mcitedefaultendpunct}{\mcitedefaultseppunct}\relax
\EndOfBibitem
\bibitem[_ig(2011)]{_igor_2011}
\emph{{IGOR} Pro v. 6.2.2.2}, 2011\relax
\mciteBstWouldAddEndPuncttrue
\mciteSetBstMidEndSepPunct{\mcitedefaultmidpunct}
{\mcitedefaultendpunct}{\mcitedefaultseppunct}\relax
\EndOfBibitem
\bibitem[Turner and Gruner(1992)]{turner_x-ray_1992}
D.~C. Turner and S.~M. Gruner, \emph{Biochemistry}, 1992, \textbf{31},
  1340{\textendash}1355\relax
\mciteBstWouldAddEndPuncttrue
\mciteSetBstMidEndSepPunct{\mcitedefaultmidpunct}
{\mcitedefaultendpunct}{\mcitedefaultseppunct}\relax
\EndOfBibitem
\bibitem[Harper \emph{et~al.}(2001)Harper, Mannock, Lewis, {McElhaney}, and
  Gruner]{harper_x-ray_2001}
P.~E. Harper, D.~A. Mannock, R.~N. Lewis, R.~N. {McElhaney} and S.~M. Gruner,
  \emph{Biophys. J.}, 2001, \textbf{81}, 2693--2706\relax
\mciteBstWouldAddEndPuncttrue
\mciteSetBstMidEndSepPunct{\mcitedefaultmidpunct}
{\mcitedefaultendpunct}{\mcitedefaultseppunct}\relax
\EndOfBibitem
\bibitem[Rappolt \emph{et~al.}(2008)Rappolt, Hodzic, Sartori, Ollivon, and
  Laggner]{rappolt_conformational_2008}
M.~Rappolt, A.~Hodzic, B.~Sartori, M.~Ollivon and P.~Laggner, \emph{Chem. Phys.
  Lipids}, 2008, \textbf{154}, 46{\textendash}55\relax
\mciteBstWouldAddEndPuncttrue
\mciteSetBstMidEndSepPunct{\mcitedefaultmidpunct}
{\mcitedefaultendpunct}{\mcitedefaultseppunct}\relax
\EndOfBibitem
\bibitem[Alley \emph{et~al.}(2008)Alley, Ces, Barahona, and
  Templer]{alley_x-ray_2008}
S.~H. Alley, O.~Ces, M.~Barahona and R.~H. Templer, \emph{Chem. Phys. Lipids},
  2008, \textbf{154}, 64--67\relax
\mciteBstWouldAddEndPuncttrue
\mciteSetBstMidEndSepPunct{\mcitedefaultmidpunct}
{\mcitedefaultendpunct}{\mcitedefaultseppunct}\relax
\EndOfBibitem
\bibitem[Ku{\v c}erka \emph{et~al.}(2006)Ku{\v c}erka, Tristram-Nagle, and
  Nagle]{kucerka_closer_2006}
N.~Ku{\v c}erka, S.~Tristram-Nagle and J.~F. Nagle, \emph{Biophys. J.}, 2006,
  \textbf{90}, L83--L85\relax
\mciteBstWouldAddEndPuncttrue
\mciteSetBstMidEndSepPunct{\mcitedefaultmidpunct}
{\mcitedefaultendpunct}{\mcitedefaultseppunct}\relax
\EndOfBibitem
\bibitem[Ku{\v c}erka \emph{et~al.}(2006)Ku{\v c}erka, Tristram-Nagle, and
  Nagle]{kucerka_structure_2006}
N.~Ku{\v c}erka, S.~Tristram-Nagle and J.~F. Nagle, \emph{J. Membr. Biol.},
  2006, \textbf{208}, 193--202\relax
\mciteBstWouldAddEndPuncttrue
\mciteSetBstMidEndSepPunct{\mcitedefaultmidpunct}
{\mcitedefaultendpunct}{\mcitedefaultseppunct}\relax
\EndOfBibitem
\bibitem[Ku{\v c}erka \emph{et~al.}(2008)Ku{\v c}erka, Nagle, Sachs, Feller,
  Pencer, Jackson, and Katsaras]{kucerka_lipid_2008}
N.~Ku{\v c}erka, J.~F. Nagle, J.~N. Sachs, S.~E. Feller, J.~Pencer, A.~Jackson
  and J.~Katsaras, \emph{Biophys. J.}, 2008, \textbf{95}, 2356--2367\relax
\mciteBstWouldAddEndPuncttrue
\mciteSetBstMidEndSepPunct{\mcitedefaultmidpunct}
{\mcitedefaultendpunct}{\mcitedefaultseppunct}\relax
\EndOfBibitem
\bibitem[Ku{\v c}erka \emph{et~al.}(2011)Ku{\v c}erka, Nieh, and
  Katsaras]{kucerka_fluid_2011}
N.~Ku{\v c}erka, M.-P. Nieh and J.~Katsaras, \emph{Biochim. Biophys. Acta,
  Biomembr.}, 2011, \textbf{1808}, 2761--2771\relax
\mciteBstWouldAddEndPuncttrue
\mciteSetBstMidEndSepPunct{\mcitedefaultmidpunct}
{\mcitedefaultendpunct}{\mcitedefaultseppunct}\relax
\EndOfBibitem
\bibitem[Kirk and Gruner(1985)]{kirk_lyotropic_1985}
G.~L. Kirk and S.~M. Gruner, \emph{Journal de Physique}, 1985, \textbf{46},
  761{\textendash}769\relax
\mciteBstWouldAddEndPuncttrue
\mciteSetBstMidEndSepPunct{\mcitedefaultmidpunct}
{\mcitedefaultendpunct}{\mcitedefaultseppunct}\relax
\EndOfBibitem
\bibitem[Vacklin \emph{et~al.}(2000)Vacklin, Khoo, Madan, Seddon, and
  Templer]{vacklin_bending_2000}
H.~Vacklin, B.~J. Khoo, K.~H. Madan, J.~M. Seddon and R.~H. Templer,
  \emph{Langmuir}, 2000, \textbf{16}, 4741--4748\relax
\mciteBstWouldAddEndPuncttrue
\mciteSetBstMidEndSepPunct{\mcitedefaultmidpunct}
{\mcitedefaultendpunct}{\mcitedefaultseppunct}\relax
\EndOfBibitem
\bibitem[Rappolt \emph{et~al.}(2003)Rappolt, Hickel, Bringezu, and
  Lohner]{rappolt_mechanism_2003}
M.~Rappolt, A.~Hickel, F.~Bringezu and K.~Lohner, \emph{Biophys. J.}, 2003,
  \textbf{84}, 3111{\textendash}3122\relax
\mciteBstWouldAddEndPuncttrue
\mciteSetBstMidEndSepPunct{\mcitedefaultmidpunct}
{\mcitedefaultendpunct}{\mcitedefaultseppunct}\relax
\EndOfBibitem
\bibitem[Kooijman \emph{et~al.}(2005)Kooijman, Chupin, Fuller, Kozlov,
  de~Kruijff, Burger, and Rand]{kooijman_spontaneous_2005}
E.~E. Kooijman, V.~Chupin, N.~L. Fuller, M.~M. Kozlov, B.~de~Kruijff, K.~N.~J.
  Burger and R.~P. Rand, \emph{Biochemistry}, 2005, \textbf{44},
  2097--2102\relax
\mciteBstWouldAddEndPuncttrue
\mciteSetBstMidEndSepPunct{\mcitedefaultmidpunct}
{\mcitedefaultendpunct}{\mcitedefaultseppunct}\relax
\EndOfBibitem
\bibitem[Safran \emph{et~al.}(1990)Safran, Pincus, and
  Andelman]{safran_theory_1990}
S.~A. Safran, P.~Pincus and D.~Andelman, \emph{Science}, 1990, \textbf{248},
  354\relax
\mciteBstWouldAddEndPuncttrue
\mciteSetBstMidEndSepPunct{\mcitedefaultmidpunct}
{\mcitedefaultendpunct}{\mcitedefaultseppunct}\relax
\EndOfBibitem
\bibitem[Kozlov and Helfrich(1992)]{kozlov_effects_1992}
M.~M. Kozlov and W.~Helfrich, \emph{Langmuir}, 1992, \textbf{8},
  2792{\textendash}2797\relax
\mciteBstWouldAddEndPuncttrue
\mciteSetBstMidEndSepPunct{\mcitedefaultmidpunct}
{\mcitedefaultendpunct}{\mcitedefaultseppunct}\relax
\EndOfBibitem
\bibitem[Keller \emph{et~al.}(1993)Keller, Bezrukov, Gruner, Tate, Vodyanoy,
  and Parsegian]{keller_probability_1993}
S.~L. Keller, S.~M. Bezrukov, S.~M. Gruner, M.~W. Tate, I.~Vodyanoy and V.~A.
  Parsegian, \emph{Biophys. J.}, 1993, \textbf{65}, 23{\textendash}27\relax
\mciteBstWouldAddEndPuncttrue
\mciteSetBstMidEndSepPunct{\mcitedefaultmidpunct}
{\mcitedefaultendpunct}{\mcitedefaultseppunct}\relax
\EndOfBibitem
\bibitem[Khelashvili \emph{et~al.}(2009)Khelashvili, Harries, and
  Weinstein]{khelashvili_modeling_2009}
G.~Khelashvili, D.~Harries and H.~Weinstein, \emph{Biophys. J.}, 2009,
  \textbf{97}, 1626--1635\relax
\mciteBstWouldAddEndPuncttrue
\mciteSetBstMidEndSepPunct{\mcitedefaultmidpunct}
{\mcitedefaultendpunct}{\mcitedefaultseppunct}\relax
\EndOfBibitem
\bibitem[May and Ben-Shaul(1995)]{may_spontaneous_1995}
S.~May and A.~Ben-Shaul, \emph{J. Chem. Phys}, 1995, \textbf{103}, 3839\relax
\mciteBstWouldAddEndPuncttrue
\mciteSetBstMidEndSepPunct{\mcitedefaultmidpunct}
{\mcitedefaultendpunct}{\mcitedefaultseppunct}\relax
\EndOfBibitem
\bibitem[Gradzielski \emph{et~al.}(1997)Gradzielski, Langevin, Sottmann, and
  Strey]{gradzielski_droplet_1997}
M.~Gradzielski, D.~Langevin, T.~Sottmann and R.~Strey, \emph{J. Chem. Phys},
  1997, \textbf{106}, 8232--8238\relax
\mciteBstWouldAddEndPuncttrue
\mciteSetBstMidEndSepPunct{\mcitedefaultmidpunct}
{\mcitedefaultendpunct}{\mcitedefaultseppunct}\relax
\EndOfBibitem
\bibitem[Koynova and Caffrey(1998)]{koynova_phases_1998}
R.~Koynova and M.~Caffrey, \emph{Biochim. Biophys. Acta, Rev. Biomembr.}, 1998,
  \textbf{1376}, 91{\textendash}145\relax
\mciteBstWouldAddEndPuncttrue
\mciteSetBstMidEndSepPunct{\mcitedefaultmidpunct}
{\mcitedefaultendpunct}{\mcitedefaultseppunct}\relax
\EndOfBibitem
\bibitem[Uppamoochikkal \emph{et~al.}(2010)Uppamoochikkal, Tristram-Nagle, and
  Nagle]{uppamoochikkal_orientation_2010}
P.~Uppamoochikkal, S.~Tristram-Nagle and J.~F. Nagle, \emph{Langmuir}, 2010,
  \textbf{26}, 17363--17368\relax
\mciteBstWouldAddEndPuncttrue
\mciteSetBstMidEndSepPunct{\mcitedefaultmidpunct}
{\mcitedefaultendpunct}{\mcitedefaultseppunct}\relax
\EndOfBibitem
\bibitem[Heberle \emph{et~al.}(2010)Heberle, Wu, Goh, Petruzielo, and
  Feigenson]{heberle_comparison_2010}
F.~A. Heberle, J.~Wu, S.~L. Goh, R.~S. Petruzielo and G.~W. Feigenson,
  \emph{Biophys. J.}, 2010, \textbf{99}, 3309--3318\relax
\mciteBstWouldAddEndPuncttrue
\mciteSetBstMidEndSepPunct{\mcitedefaultmidpunct}
{\mcitedefaultendpunct}{\mcitedefaultseppunct}\relax
\EndOfBibitem
\bibitem[Ionova \emph{et~al.}(2012)Ionova, Livshits, and
  Marsh]{ionova_phase_2012}
I.~V. Ionova, V.~A. Livshits and D.~Marsh, \emph{Biophys. J.}, 2012,
  \textbf{102}, 1856--1865\relax
\mciteBstWouldAddEndPuncttrue
\mciteSetBstMidEndSepPunct{\mcitedefaultmidpunct}
{\mcitedefaultendpunct}{\mcitedefaultseppunct}\relax
\EndOfBibitem
\bibitem[Lingwood and Simons(2010)]{lingwood_lipid_2010}
D.~Lingwood and K.~Simons, \emph{Science}, 2010, \textbf{327},
  46{\textendash}50\relax
\mciteBstWouldAddEndPuncttrue
\mciteSetBstMidEndSepPunct{\mcitedefaultmidpunct}
{\mcitedefaultendpunct}{\mcitedefaultseppunct}\relax
\EndOfBibitem
\bibitem[Smith and Freed(2009)]{smith_determination_2009}
A.~K. Smith and J.~H. Freed, \emph{J. Phys. Chem. B}, 2009, \textbf{113},
  3957--3971\relax
\mciteBstWouldAddEndPuncttrue
\mciteSetBstMidEndSepPunct{\mcitedefaultmidpunct}
{\mcitedefaultendpunct}{\mcitedefaultseppunct}\relax
\EndOfBibitem
\bibitem[Ben-Shaul(1995)]{ben-shaul_molecular_1995}
A.~Ben-Shaul, \emph{Handbook of biological physics}, North-Holland, 1995,
  vol.~1, p. 359{\textendash}401\relax
\mciteBstWouldAddEndPuncttrue
\mciteSetBstMidEndSepPunct{\mcitedefaultmidpunct}
{\mcitedefaultendpunct}{\mcitedefaultseppunct}\relax
\EndOfBibitem
\bibitem[Sch{\"a}fer \emph{et~al.}(2011)Sch{\"a}fer, de~Jong, Holt, Rzepiela,
  de~Vries, Poolman, Killian, and Marrink]{schafer_lipid_2011}
L.~V. Sch{\"a}fer, D.~H. de~Jong, A.~Holt, A.~J. Rzepiela, A.~H. de~Vries,
  B.~Poolman, J.~A. Killian and S.~J. Marrink, \emph{{PNAS}}, 2011,
  \textbf{108}, 1343{\textendash}1348\relax
\mciteBstWouldAddEndPuncttrue
\mciteSetBstMidEndSepPunct{\mcitedefaultmidpunct}
{\mcitedefaultendpunct}{\mcitedefaultseppunct}\relax
\EndOfBibitem
\bibitem[Doma{\'n}ski \emph{et~al.}(2012)Doma{\'n}ski, Marrink, and
  Sch{\"a}fer]{domanski_transmembrane_2012}
J.~Doma{\'n}ski, S.~J. Marrink and L.~V. Sch{\"a}fer, \emph{Biochim. Biophys.
  Acta, Biomembr.}, 2012, \textbf{1818}, 984{\textendash}994\relax
\mciteBstWouldAddEndPuncttrue
\mciteSetBstMidEndSepPunct{\mcitedefaultmidpunct}
{\mcitedefaultendpunct}{\mcitedefaultseppunct}\relax
\EndOfBibitem
\bibitem[Garc{\'i}a-S{\'a}ez \emph{et~al.}(2007)Garc{\'i}a-S{\'a}ez, Chiantia,
  and Schwille]{garcia-saez_effect_2007}
A.~J. Garc{\'i}a-S{\'a}ez, S.~Chiantia and P.~Schwille, \emph{J. Biol. Chem.},
  2007, \textbf{282}, 33537--33544\relax
\mciteBstWouldAddEndPuncttrue
\mciteSetBstMidEndSepPunct{\mcitedefaultmidpunct}
{\mcitedefaultendpunct}{\mcitedefaultseppunct}\relax
\EndOfBibitem
\bibitem[Lee \emph{et~al.}(2011)Lee, Min, Dhar, Ramachandran, Israelachvili,
  and Zasadzinski]{lee_relating_2011}
D.~W. Lee, Y.~Min, P.~Dhar, A.~Ramachandran, J.~N. Israelachvili and J.~A.
  Zasadzinski, \emph{{PNAS}}, 2011, \textbf{108}, 9425--9430\relax
\mciteBstWouldAddEndPuncttrue
\mciteSetBstMidEndSepPunct{\mcitedefaultmidpunct}
{\mcitedefaultendpunct}{\mcitedefaultseppunct}\relax
\EndOfBibitem
\bibitem[Helfrich(1973)]{helfrich_elastic_1973}
W.~Helfrich, \emph{Z. Naturforsch., C: J. Biosci.}, 1973,  693---703\relax
\mciteBstWouldAddEndPuncttrue
\mciteSetBstMidEndSepPunct{\mcitedefaultmidpunct}
{\mcitedefaultendpunct}{\mcitedefaultseppunct}\relax
\EndOfBibitem
\bibitem[Khelashvili \emph{et~al.}(2013)Khelashvili, Kollmitzer, Heftberger,
  Pabst, and Harries]{khelashvili_calculating_2013}
G.~Khelashvili, B.~Kollmitzer, P.~Heftberger, G.~Pabst and D.~Harries, \emph{J.
  Chem. Theory Comput.}, 2013\relax
\mciteBstWouldAddEndPuncttrue
\mciteSetBstMidEndSepPunct{\mcitedefaultmidpunct}
{\mcitedefaultendpunct}{\mcitedefaultseppunct}\relax
\EndOfBibitem
\bibitem[Risselada and Marrink(2008)]{risselada_molecular_2008}
H.~J. Risselada and S.~J. Marrink, \emph{{PNAS}}, 2008, \textbf{105},
  17367--17372\relax
\mciteBstWouldAddEndPuncttrue
\mciteSetBstMidEndSepPunct{\mcitedefaultmidpunct}
{\mcitedefaultendpunct}{\mcitedefaultseppunct}\relax
\EndOfBibitem
\bibitem[Tian \emph{et~al.}(2007)Tian, Johnson, Wang, and
  Baumgart]{tian_line_2007}
A.~Tian, C.~Johnson, W.~Wang and T.~Baumgart, \emph{Phys. Rev. Lett.}, 2007,
  \textbf{98}, 208102\relax
\mciteBstWouldAddEndPuncttrue
\mciteSetBstMidEndSepPunct{\mcitedefaultmidpunct}
{\mcitedefaultendpunct}{\mcitedefaultseppunct}\relax
\EndOfBibitem
\bibitem[Esposito \emph{et~al.}(2007)Esposito, Tian, Melamed, Johnson, Tee, and
  Baumgart]{esposito_flicker_2007}
C.~Esposito, A.~Tian, S.~Melamed, C.~Johnson, S.-Y. Tee and T.~Baumgart,
  \emph{Biophysical Journal}, 2007, \textbf{93}, 3169--3181\relax
\mciteBstWouldAddEndPuncttrue
\mciteSetBstMidEndSepPunct{\mcitedefaultmidpunct}
{\mcitedefaultendpunct}{\mcitedefaultseppunct}\relax
\EndOfBibitem
\bibitem[Honerkamp-Smith \emph{et~al.}(2008)Honerkamp-Smith, Cicuta, Collins,
  Veatch, den Nijs, Schick, and Keller]{honerkamp-smith_line_2008}
A.~R. Honerkamp-Smith, P.~Cicuta, M.~D. Collins, S.~L. Veatch, M.~den Nijs,
  M.~Schick and S.~L. Keller, \emph{Biophysical Journal}, 2008, \textbf{95},
  236--246\relax
\mciteBstWouldAddEndPuncttrue
\mciteSetBstMidEndSepPunct{\mcitedefaultmidpunct}
{\mcitedefaultendpunct}{\mcitedefaultseppunct}\relax
\EndOfBibitem
\bibitem[Heberle \emph{et~al.}(2013)Heberle, Petruzielo, Pan, Drazba, Ku{\v
  c}erka, Standaert, Feigenson, and Katsaras]{heberle_bilayer_2013}
F.~A. Heberle, R.~S. Petruzielo, J.~Pan, P.~Drazba, N.~Ku{\v c}erka, R.~F.
  Standaert, G.~W. Feigenson and J.~Katsaras, \emph{J. Am. Chem. Soc.}, 2013,
  \textbf{135}, 6853--6859\relax
\mciteBstWouldAddEndPuncttrue
\mciteSetBstMidEndSepPunct{\mcitedefaultmidpunct}
{\mcitedefaultendpunct}{\mcitedefaultseppunct}\relax
\EndOfBibitem
\end{mcitethebibliography}
\providecommand*{\mcitethebibliography}{\thebibliography}
\csname @ifundefined\endcsname{endmcitethebibliography}
{\let\endmcitethebibliography\endthebibliography}{}

}

\end{document}